\documentclass[runningheads]{llncs}
\usepackage[T1]{fontenc}
\usepackage{graphicx}
\usepackage{url}
\usepackage{hyperref}
\usepackage{amsmath,amssymb}
\usepackage{isabelle,isabellesym}
\usepackage{color}

\newcommand{\card}[1]{\lvert#1\rvert}
\newcommand{\omitted}{\textrm{\emph omitted}}
\begin{document}
\title{A Formalisation of a Special Case of the Union-Closed Conjecture in Isabelle/HOL}
\titlerunning{Formalisation of a special case of the Union-Closed Conjecture}
\author{
Angeliki Koutsoukou-Argyraki \inst{1,2}\orcidID{0000-0002-8886-5281} \and \newline
Lawrence C. Paulson \inst{2}\orcidID{0000-0003-0288-4279 }}
\authorrunning{Koutsoukou-Argyraki, A. \and Paulson, L. C.}
%
\institute{Royal Holloway, University of London, Egham, Surrey TW20 0EX, UK \and
University of Cambridge, 15 JJ Thomson Avenue, Cambridge
CB3 0FD, UK \\
\email{angeliki.koutsoukouargyraki@rhul.ac.uk, lp15@cam.ac.uk}\\
 }
\maketitle              
\begin{abstract} A 2021 proof of a special case of the Union-Closed Conjecture, by Aaronson, Ellis and Leader, 
has been formalised in the proof assistant Isabelle/HOL\@. 
Our discussion involves sketching their proof and displaying snippets from the Isabelle version of the proof, illustrating the extent to which mathematical reasoning can be rendered clearly in a formal language.

\keywords{Union Closed Conjecture \and additive combinatorics \and number theory \and Isabelle/HOL \and interactive theorem proving \and formalisation of mathematics.}
\end{abstract}

\section{Introduction}
Given a set $X$, a family $\mathcal{F}$  of subsets of $X$ is said to be \textit{union-closed} if the union of any two sets from $\mathcal{F}$ is also in $\mathcal{F}$. A celebrated 1979 conjecture, attributed to Frankl,
states that for every union-closed family $\mathcal{F}$ of subsets of a finite set $X$ (with $\mathcal{F} \neq \{\emptyset\}$), there exists an element $x \in X$ that belongs to at least half of the sets in~$\mathcal{F}$~\cite{Aaronson_et_al_union_closed}.
For example, the family $\emptyset, \{a\}, \{a, b\}, \{ b, c, d, e\}, \{a, b, c, d, e\}$ is union-closed.
The conjecture holds in this case because element $a$ is in 3 out of the 5 sets of the family ($b$ also works).

The conjecture has been shown for several special cases of union-closed families or where constraints are imposed on the set~$X$.
Aaronson et al. \cite{Aaronson_et_al_union_closed} list a number of such results.
The proof of the general case remains elusive despite the simplicity of the statement. In the winter 2022--23 there was a renewal of interest because of a significant 
breakthrough by Gilmer, who gave the first constant lower bound for
the proportion of the most common element in a union-closed family \cite{gilmer}.
This result fuelled a number of related works towards the conjecture by Alweiss, Huang and Sellke \cite{Alweiss_et_al}, Chase and Lovett \cite{Chase_Lovett}, Sawin \cite{sawin} and others. 
Cambie summarizes the developments made in the winter of 2022–2023 in a comprehensive review
\cite{Cambie_review}. Despite these promising developments, the general case remains unsolved.

In this paper we discuss our formalisation \cite{Transitive_Union_Closed_Families-AFP} in the proof assistant Isabelle/HOL of a proof of a special case of the Union-Closed Conjecture that had been shown by Aaronson, Ellis and Leader in 2021~\cite{Aaronson_et_al_union_closed}, namely that the conjecture holds for the union-closed family generated by the cyclic translates of any fixed set:

\begin{theorem} \label{main} \textbf{(Aaronson, Ellis and Leader, 2021)}: 
Let $ (G, +)$ be a finite Abelian group, and let $R \subseteq G$ with $R \neq \emptyset$. Let $\mathcal{F} = \{ A+ R:~ A \subseteq G \}$ be the set of all unions of translates of $R$. Then the average size of a set in $\mathcal{F}$ is at least $|G|/2$. In particular, the Union-Closed Conjecture holds for $\mathcal{F}$.
\end{theorem}

The proof is elegant and surprisingly short (one page in the original paper \cite{Aaronson_et_al_union_closed}). The Isabelle/HOL formalisation \cite{Transitive_Union_Closed_Families-AFP} is equally brief, consisting of only about 270 lines of code.
We adopt some techniques from additive combinatorics: The main essential prerequisites involve sumset theory. Given $A$, $B$ finite subsets of an Abelian group, the \textit{sumset} $A+ B$ is defined as the set $\{a + b : a \in A,\allowbreak b \in B \}$. 

We had previously developed sumset theory in an AFP entry 
formalising the Pl\"{u}nnecke–Ruzsa Inequality \cite{Pluennecke_Ruzsa_Inequality-AFP}, 
which in turn depended on Ballarin's algebra development \cite{Jacobson_Basic_Algebra-AFP}. Other Isabelle proof developments building on sumset theory include theorems of central importance such as
the Balog--Szemer\'{e}di--Gowers Theorem, 
\cite{Balog_Szemeredi_Gowers-AFP,Balog_Szemeredi_Gowers_paper},
Khovanskii's Theorem
\cite{Khovanskii_Theorem-AFP}, Kneser's Theorem and the Cauchy--Davenport Theorem
\cite{Kneser_Cauchy_Davenport-AFP}
\footnote{For a summary see the first author's invited talk at the 14th International Conference on Interactive Theorem Proving \cite{koutsoukou_itp_2023}.}, as well as a later generalisation of the Cauchy--Davenport Theorem \cite{Generalized_Cauchy_Davenport-AFP}.

The first author discussed this formalisation \cite{Transitive_Union_Closed_Families-AFP} during an invited talk at the 2025 Big Proof workshop, at the Isaac Newton Institute for Mathematical Sciences in Cambridge:\footnote{\url{https://www.newton.ac.uk/event/bprw03/}}
 \textit{A Formalisation of a Special Case of the Union-Closed Conjecture and Formal Proofs as Datasets in the Era of LLMs}.
 A primary focus at that event was \textit{autoformalisation}, using the neurosymbolic tools that have been increasingly taking over mathematical practice.
 A discussion of this formalisation was thus used to explore different possible technical formalisation choices that a human formaliser
 can make, considering certain aspects of the proof, when thinking of formal proofs as potential training data for LLMs.
Human-written formal proofs are used as training data for AI tools, so we want them to be of high quality, which includes being legible to AI tools, in addition to human readers as we hope.
No LLMs were used in the preparation of our formalisation \cite{Transitive_Union_Closed_Families-AFP}.

In Section~2 we give a proof sketch and discuss the formalisation. Section~3 briefly presents the formalisation of some arguments not explicit in the original paper, and we conclude in Section~4.

\section{A formal proof of the main theorem}

We interleave a sketch of Aaronson et al.'s proof with fragments of our Isabelle formalisation.
The purpose of including the formal material is to convey an impression 
of how the mathematics can be communicated both to the machine and to human readers.
We want our proofs to be informative as well as formal.

\subsection{Preliminaries}

A key definition is that of a union-closed family of sets:
\begin{isabelle}
\isakeywordONE{definition}\ union\_closed::\ "'a\ set\ set\ \isasymRightarrow \ bool"\ \isanewline
\ \ \isakeywordTWO{where}\ "union\_closed\ \isasymF \ \isasymequiv \ (\isasymforall A\isasymin \isasymF .\ \isasymforall B\isasymin \isasymF .\ A\isasymunion B\ \isasymin \ \isasymF )"
\end{isabelle}

Next we define the context for the main theorem, not an arbitrary union-closed family but one generated in a particular way.
For this, we use an Isabelle \textit{locale}:
\begin{isabelle}
\isakeywordONE{locale}\ Family\ =\ additive\_abelian\_group\ +\isanewline
\ \ \isakeywordTWO{fixes}\ R\ \isanewline
\ \ \isakeywordTWO{assumes}\ finG:\ "finite\ G"\isanewline
\ \ \isakeywordTWO{assumes}\ RG:\ "R\ \isasymsubseteq \ G"\isanewline
\ \ \isakeywordTWO{assumes}\ R\_nonempty:\ "R\ \isasymnoteq \ \{\}"
\end{isabelle}
A locale is essentially a portable context, where variables and their properties can be declared.
Working within a locale implicitly imports those variables and properties.
Here we extend an existing locale for an additive Abelian group~$G$ with an assumption of finiteness and a fixed, non-empty subset~$R$.

Within the locale we can refer to $R$. Given $A\subseteq G$, the \textit{neighbourhood} of $A$ is defined to be the sumset $A+R$.
Now $\mathcal{F}$ is the set of all neighbourhoods generated by all subsets of G. 
(The operator \isa{`} denotes image, while \isa{Pow} denotes powerset.)
\begin{isabelle}
\isakeywordONE{definition}\ "Neighbd\ \isasymequiv \ \isasymlambda A.\ sumset\ A\ R"\isanewline
\isakeywordONE{definition}\ "\isasymF \ \isasymequiv \ Neighbd\ `\ Pow\ G"
\end{isabelle}

The complete statement of Theorem \ref{main} in our formalisation \cite{Transitive_Union_Closed_Families-AFP} is written in Isabelle as follows, considering first the lemma:
\begin{isabelle}
\isakeywordONE{lemma}\ average\_ge:\ \isanewline
\ \ \isakeywordTWO{shows}\ "(\isasymSum S\isasymin \isasymF .(card\ S))\ /\ card\ \isasymF \ \isasymge \ card\ G\ /\ 2"
\end{isabelle}
This lemma straightforwardly implies the union-closed conjecture for the family~$\mathcal{F}$.
We define the union-closed property itself: 
\begin{isabelle}
\isakeywordONE{definition}\ union\_closed\_conjecture\_property::\ "'a\ set\ set\ \isasymRightarrow \ bool"\isanewline
\ \ \isakeywordTWO{where}\ "union\_closed\_conjecture\_property\ \isasymF \ \isanewline
\ \ \ \ \ \ \ \isasymequiv \ \isasymexists \isasymX \isasymsubseteq \isasymF .\ \isasymexists x\isasymin G.\ x\ \isasymin \ \isasymInter \isasymX \ \isasymand \ card\ \isasymX \ \isasymge \ card\ \isasymF \ /\ 2"
\end{isabelle}
In Section~3, by a simple counting argument, we will eventually prove the desired conclusion: 
\begin{isabelle}
\isakeywordONE{theorem}\ Aaronson\_Ellis\_Leader\_union\_closed\_conjecture:\isanewline
\ \ \isakeywordTWO{shows}\ "union\_closed\_conjecture\_property\ \isasymF "
\end{isabelle}

Note that the family $\mathcal{F}$ is indeed union-closed.
For any $S_1$, $S_2 \in \mathcal{F}$ we obtain $A_1$, $A_2 \subseteq G$ and need to show that 
$(A_1 + R) \cup (A_2 + R) = (A_1 \cup A_2) + R \in \mathcal{F}$.
In Isabelle, that is straightforward to show by applying elementary 
lemmas of sumset theory with basic automation:
\begin{isabelle}
\isakeywordONE{lemma}\ "union\_closed\ \isasymF "\ \ \ \ \isanewline
\isakeywordONE{proof} -\isanewline
\ \ \isakeywordONE{have}\ "\isasymforall A\isasymsubseteq G.\ \isasymforall B\isasymsubseteq G.\ (sumset\ A\ R)\ \isasymunion \ (sumset\ B\ R)\ =\ sumset\ (A\ \isasymunion \ B)\ R"\isanewline
\ \ \ \ \isakeywordONE{by}\ (simp\ add:\ sumset\_subset\_Un1)\isanewline
\ \ \isakeywordTHREE{then show}\ ?thesis\isanewline
\ \ \ \ \isakeywordONE{by}\ (auto\ simp:\ union\_closed\_def\ \isasymF \_def\ Neighbd\_def)\isanewline
\isakeywordONE{qed}%
\end{isabelle}

The \textbf{by} command gives the proof justification, which we will not discuss in any detail.
Typical justifications are by simplification, by logical reasoning or a combination of the two.

\subsection{Proof sketch}

For any set $A \subseteq G$
its \emph{$R$-interior} is defined to be
$$\text{Int}_R(A) := \{ x \in G~:~ x+R \subseteq A\}.$$

Then we define a function $f: \mathcal{P}(G) \rightarrow 
\mathcal{P}(G)$  by $f(S) =- ( G \setminus \text{Int}_R (S))$
for all $S \subseteq G$.
The key step of the theorem is to show that $f\restriction{\mathcal{F}}$
is a bijection from $\mathcal{F}$ to itself and satisfies 
\begin{equation} \label{eqn:one}
\card{S} + |f(S)|  \geq |G|\quad \text{for all $S \in \mathcal{F}.$}
\end{equation}

It follows by a fairly straightforward calculation that
\begin{equation} \label{eqn:two}
f(S) = (-(G \setminus S)) + R   \quad \text{for all $S \subseteq G$}, 
\end{equation}
which implies that $f(\mathcal{P}(G)) \subseteq \mathcal{F}$. Moreover,

\begin{equation} \label{eqn:three}
N_R(\text{Int}_R(A + R)) = A + R
\end{equation}
Thus $f\restriction\mathcal{F}$ is a bijection from $\mathcal{F}$ to itself.
That implies $\sum_{S \in \mathcal{F}} \card{f(S)} = \sum_{S \in \mathcal{F}} \card{S}$,
from which we compute
\begin{equation} \label{eqn:final}
\frac{1}{\card{\mathcal{F}}} \sum_{S \in \mathcal{F}} \card{S} =
\frac{1}{2 \card{\mathcal{F}}} \sum_{S \in \mathcal{F}} 
( \card{S} + \card{f(S)}) \geq
\frac{1}{2 \card{\mathcal{F}}} \sum_{S \in \mathcal{F}}  |G| = |G|/2.
\end{equation}

We have thus shown that the average size of a set in the family $\mathcal{F}$ is at
least $|G|/2$, so the first part of Theorem \ref{main} is complete.
As this is the theorem proved by Aaronson et al.~\cite{Aaronson_et_al_union_closed}, let's turn to its formalisation.
First, a definition:

\begin{isabelle}
\isakeywordONE{definition}\ "Interior\ \isasymequiv \ \isasymlambda A.\ \{x\isasymin G.\ sumset\ \{x\}\ R\ \isasymsubseteq \ A\}"
\end{isabelle}

A useful fact is that $\card{\text{Int}_R(A)} \le \card S$ for $S\subseteq G$.
It holds because the operation
of addition by any fixed $r\in R$ is an injection from $\text{Int}_R(A)$ into~$S$. The argument should be discernible in the formal proof below,
noting that \isa{inj\_on~f~T} expresses that \isa{f} is injective on the set~\isa{T} and that
\isa{f ‘ T} is the image of \isa{T} under~\isa{f}.

\begin{isabelle}
\isakeywordONE{lemma}\ card\_Interior\_le:\isanewline
\ \ \isakeywordTWO{assumes}\ "S\ \isasymsubseteq \ G"\isanewline
\ \ \isakeywordTWO{shows}\ "card\ (Interior\ S)\ \isasymle \ card\ S"\isanewline
\isakeywordONE{proof}\ -\isanewline
\ \ \isakeywordTHREE{obtain}\ r\ \isakeywordTWO{where}\ "r\ \isasymin \ R"\isanewline
\ \ \ \ \isakeywordONE{using}\ R\_nonempty\ \isakeywordONE{by}\ blast\isanewline
\ \ \isakeywordTHREE{show}\ ?thesis\isanewline
\ \ \isakeywordONE{proof}\ (intro\ card\_inj\_on\_le)\isanewline
\ \ \ \ \isakeywordONE{let}\ ?f\ =\ "(\isasymlambda x.\ x\ \isasymoplus \ r)"\isanewline
\ \ \ \ \isakeywordTHREE{show}\ "inj\_on\ ?f\ (Interior\ S)"\ "?f\ `\ Interior\ S\ \isasymsubseteq \ S"\isanewline
\ \ \ \ \ \ \isakeywordONE{using}\ RG\ \isacartoucheopen r\ \isasymin \ R\isacartoucheclose \ \isakeywordONE{by}\ (auto\ simp:\ Interior\_def\ inj\_on\_def)\isanewline
\ \ \ \ \isakeywordTHREE{show}\ "finite\ S"\isanewline
\ \ \ \ \ \ \isakeywordONE{using}\ assms\ finG\ finite\_subset\ \isakeywordONE{by}\ blast\isanewline
\ \ \isakeywordONE{qed}\isanewline
\isakeywordONE{qed}%
\end{isabelle}

Let's walk through the formal proof of the main theorem step-by-step. 
First, here is the theorem statement:

\begin{isabelle}
\isakeywordONE{lemma}\ average\_ge:\ \isanewline
\ \ \isakeywordTWO{shows}\ "(\isasymSum S\isasymin \isasymF .(card\ S))\ /\ card\ \isasymF \ \isasymge \ card\ G\ /\ 2"\isanewline
\isakeywordONE{proof}-
\end{isabelle}

\noindent
Following the development above, we define the function $f$ and prove equation~(\ref{eqn:one}).
The omitted proof is by elementary combinatorics and the theorem \isa{card\_Interior\_le}.
\begin{isabelle}
\ \ \isakeywordTHREE{define}\ f\ \isakeywordTWO{where}\ "f\ \isasymequiv \ \isasymlambda S.\ minusset\ (G\ \isasymsetminus \ Interior\ S)"\isanewline
\ \ \isakeywordONE{have}\ 1:\ "card\ S\ +\ card\ (f\ S)\ \isasymge \ card\ G"\ \isakeywordTWO{if}\ "S\ \isasymsubseteq \ G"\ \isakeywordTWO{for}\ S\isanewline
\ \ \isakeywordONE{proof}\ - \omitted\ \isakeywordONE{qed}%
\end{isabelle}

\noindent
The proof of equation~(\ref{eqn:two}) begins as follows:
\begin{isabelle}
\ \ \isakeywordONE{have}\ 2:\ "f\ S\ =\ sumset\ (minusset\ (G\ \isasymsetminus \ S))\ R"\ \isakeywordTWO{if}\ "S\ \isasymsubseteq \ G"\ \isakeywordTWO{for}\ S\isanewline
\ \ \isakeywordONE{proof}\ -
\end{isabelle}

\noindent
It relies on a fact proved by a series of logical equivalences.
Note that the symbol \isa{...}, which is a literal part of the Isabelle proof text, refers to the previous right-hand side.
\begin{isabelle}
\ \ \ \ \isakeywordONE{have}\ *:\ "x\ \isasymin \ f\ S\ \isasymlongleftrightarrow \ x\ \isasymin \ sumset\ (minusset(G\isasymsetminus S))\ R"\ \isakeywordTWO{if}\ "x\ \isasymin \ G"\ \isakeywordTWO{for}\ x\isanewline
\ \ \ \ \isakeywordONE{proof}\ -\isanewline
\ \ \ \ \ \ \isakeywordONE{have}\ "x\ \isasymin \ f\ S\ \isasymlongleftrightarrow \ inverse\ x\ \isasymnotin \ Interior\ S"\isanewline
\ \ \ \ \ \ \ \ \isakeywordONE{using}\ that\ minusset.simps\ \isakeywordONE{by}\ (fastforce\ simp:\ f\_def)+\isanewline
\ \ \ \ \ \ \isakeywordONE{also}\ \isakeywordONE{have}\ "\isasymdots \ \isasymlongleftrightarrow \ (sumset\ \{inverse\ x\}\ R)\ \isasyminter \ (G\isasymsetminus S)\ \isasymnoteq \ \{\}"\isanewline
\ \ \ \ \ \ \ \ \isakeywordONE{using}\ sumset\_subset\_carrier\ that\ \isakeywordONE{by}\ (auto\ simp:\ Interior\_def)\isanewline
\ \ \ \ \ \ \isakeywordONE{also}\ \isakeywordONE{have}\ "\isasymdots \ \isasymlongleftrightarrow \ x\ \isasymin \ sumset\ (minusset\ (G\isasymsetminus S))\ R"\isanewline
\ \ \ \ \ \ \isakeywordONE{proof}\ {\omitted}\ \isakeywordONE{qed}\isanewline
\ \ \ \ \ \ \isakeywordONE{finally}\ \isakeywordTHREE{show}\ ?thesis\ \isakeywordONE{.}\isanewline
\ \ \ \ \isakeywordONE{qed}
\end{isabelle}

\noindent
With fact (\isa{*}) proved, equation~(\ref{eqn:two}) has a separate proof in each direction.
\begin{isabelle}
\ \ \ \ \isakeywordTHREE{show}\ ?thesis\isanewline
\ \ \ \ \isakeywordONE{proof}\isanewline
\ \ \ \ \ \ \isakeywordTHREE{show}\ "f\ S\ \isasymsubseteq \ sumset\ (minusset\ (G\ \isasymsetminus \ S))\ R"\isanewline
\ \ \ \ \ \ \isakeywordONE{using}\ "*"\ f\_def\ minusset\_subset\_carrier\ \isakeywordONE{by}\ blast\isanewline
\ \ \ \ \isakeywordONE{next}\isanewline
\ \ \ \ \ \ \isakeywordTHREE{show}\ "sumset\ (minusset\ (G\ \isasymsetminus \ S))\ R\ \isasymsubseteq \ f\ S"\isanewline
\ \ \ \ \ \ \isakeywordONE{by}\ (meson\ "*"\ subset\_iff\ sumset\_subset\_carrier)\isanewline
\ \ \ \ \isakeywordONE{qed}\isanewline
\ \ \isakeywordONE{qed}
\end{isabelle}

\noindent
Then we immediately prove the claimed consequence of~(\ref{eqn:two}).
\begin{isabelle}
\ \ \isakeywordONE{then}\ \isakeywordONE{have}\ "f\ `\ Pow\ G\ \isasymsubseteq \ \isasymF "\isanewline
\ \ \ \ \isakeywordONE{by}\ (auto\ simp:\ Neighbd\_def\ \isasymF \_def\ minusset\_subset\_carrier)
\end{isabelle}

\noindent
The key fact~(\ref{eqn:three}) is proved automatically from the definitions.
\begin{isabelle}
\ \ \isakeywordONE{have}\ 3:\ "Neighbd\ (Interior\ (sumset\ A\ R))\ =\ sumset\ A\ R"\ \isanewline
\ \ \ \ \isakeywordTWO{if}\ "A\ \isasymsubseteq \ G"\ \isakeywordTWO{for}\ A\isanewline
\ \ \ \ \isakeywordONE{using}\ that\ \isakeywordONE{by}\ (force\ simp:\ sumset\_eq\ Neighbd\_def\ Interior\_def)
\end{isabelle}

\noindent
Skipping some routine calculations, the formal proof ends as follows:
\begin{isabelle}
\ \ \isakeywordONE{have}\ "card\ G\ /\ 2\ =\ (1\ /\ (2\ *\ card\ \isasymF ))\ *\ (\isasymSum S\isasymin \isasymF .\ card\ G)"\isanewline
\ \ \ \ \isakeywordONE{by}\ simp\isanewline
\ \ \isakeywordONE{also}\ \isakeywordONE{have}\ "\isasymdots \ \isasymle \ (1\ /\ (2\ *\ card\ \isasymF ))\ *\ (\isasymSum S\isasymin \isasymF .\ card\ S\ +\ card\ (f\ S))"\isanewline
\ \ \ \ \isakeywordONE{by}\ (intro\ sum\_mono\ mult\_left\_mono\ of\_nat\_mono\ 1)\ (auto\ simp:\ \isasymF \_def)\isanewline
\ \ \isakeywordONE{also}\ \isakeywordONE{have}\ "\isasymdots \ =\ (1\ /\ card\ \isasymF )\ *\ (\isasymSum S\isasymin \isasymF .\ card\ S)"\isanewline
\ \ \ \ \isakeywordONE{by}\ (simp\ add:\ sum\_card\_eq\ sum.distrib)\isanewline
\ \ \isakeywordONE{finally}\ \isakeywordTHREE{show}\ ?thesis\isanewline
\ \ \ \ \isakeywordONE{by}\ argo\isanewline
\isakeywordONE{qed}%
\end{isabelle}

The formal proof of the main theorem, including the parts omitted above, is short: 100 lines of Isabelle/HOL\@.

\section{Proving that union-closed conjecture holds for $\mathcal{F}$} \label{sec:uc-holds}

The authors' theorem straightforwardly implies the union-closed conjecture for the family~$\mathcal{F}$,
but they did not give the details~\cite{Aaronson_et_al_union_closed}.
For avoidance of all doubt, we decided to formalise this last bit of reasoning.

A key argument, which required about 28 lines in the formal proof,
was to count the elements of~$G$ in two different ways:%
\footnote{A typo in the original paper \cite{Aaronson_et_al_union_closed} reads $x \in \mathcal{F}$ instead of 
the obvious $x \in S$.}
\begin{equation} \label{eqn:sumeq}
\sum_{x \in G} \card{\{ S \in \mathcal{F}~:~ x \in S \}} = \sum_{S \in \mathcal{F}} \card{S}.
\end{equation}
Each summand on the left is the number of occurrences of a given~$x\in G$ among the elements of $\mathcal F$.
The proof is straightforward, by induction on~$\mathcal{F}$.

From the above and the first part of the theorem~(\ref{eqn:final}) we derive
\begin{equation} \label{eqn:F2}
\frac{1}{\card{G}} \sum_{x \in G} \card{\{ S \in \mathcal{F}~:~ x \in S \}} \geq
\frac{\card{\mathcal{F}}}{2},  
\end{equation}
which shows that the Union-Closed Conjecture holds for $\mathcal{F}$.
Otherwise every set $\{ S \in \mathcal{F}~:~ x \in S \}$
would have to be smaller than $\card{\mathcal{F}}/2$ and therefore the average too.
This argument is about 25 lines in the formal proof \cite{Transitive_Union_Closed_Families-AFP}.

Now let us look at the formalisation of this final part.
First, the theorem statement:
\begin{isabelle}
\isakeywordONE{theorem}\ Aaronson\_Ellis\_Leader\_union\_closed\_conjecture:\isanewline
\ \ \isakeywordTWO{shows}\ "union\_closed\_conjecture\_property\ \isasymF "\isanewline
\isakeywordONE{proof}\ -
\end{isabelle}

Now let us begin the proof of the key identity~(\ref{eqn:sumeq}). The base case of the induction is trivial.
\begin{isabelle}
\isakeywordONE{have}\ *:\ "(\isasymSum S\isasymin \isasymF .(card\ S))\ =\ (\isasymSum x\isasymin G.\ card\ \{S\isasymin \isasymF .\ x\isasymin S\})"\isanewline
\ \ \ \ \isakeywordONE{using}\ finite\isasymF \ \isasymF \_subset\isanewline
\ \ \isakeywordONE{proof}\ induction\isanewline
\ \ \ \ \isakeywordTHREE{case}\ empty\isanewline
\ \ \ \ \isakeywordONE{then}\ \isakeywordTHREE{show}\ ?case\ \isanewline
\ \ \ \ \ \ \isakeywordONE{by}\ simp\isanewline
\ \ \isakeywordONE{next}
\end{isabelle}

The inductive argument for a new \isa{S} in some $\mathcal{G}$ (a subset of $\mathcal{F}$) 
distributes and counts the elements of \isa{S} as shown below:
\begin{isabelle}
\ \ \ \ \isakeywordTHREE{case}\ (insert\ S\ \isasymG )\isanewline
\ \ \ \ \isakeywordONE{then}\ \isakeywordONE{have}\ A:\ "\{T.\ (T\ =\ S\ \isasymor \ T\isasymin \isasymG )\ \isasymand \ x\isasymin T\}\ \isanewline
\ \ \ \ \ \ \ \ \ \ \ \ \ \ \ \ \ =\ \{T\isasymin \isasymG .\ x\isasymin T\}\ \isasymunion \ (if\ x\isasymin S\ then\ \{S\}\ else\ \{\})"\isanewline
\ \ \ \ \ \ \isakeywordTWO{for}\ x\isanewline
\ \ \ \ \ \ \isakeywordONE{by}\ auto\isanewline
\ \ \ \ \isakeywordONE{have}\ B:\ "card\ \{T.\ (T\ =\ S\ \isasymor \ T\isasymin \isasymG )\ \isasymand \ x\isasymin T\}\ \isanewline
\ \ \ \ \ \ \ \ \ \ \ =\ card\ \{T\isasymin \isasymG .\ x\isasymin T\}\ +\ (if\ x\isasymin S\ then\ 1\ else\ 0)"\isanewline
\ \ \ \ \ \ \isakeywordTWO{for}\ x\isanewline
\ \ \ \ \ \ \isakeywordONE{by}\ (simp\ add:\ A\ card\_insert\_if\ insert)\isanewline
\ \ \ \ \isakeywordONE{have}\ "S\ =\ (\isasymUnion x\isasymin G.\ if\ x\ \isasymin \ S\ then\ \{x\}\ else\ \{\})"\isanewline
\ \ \ \ \ \ \isakeywordONE{using}\ insert.prems\ \isakeywordONE{by}\ auto\isanewline
\ \ \ \ \isakeywordONE{then}\ \isakeywordONE{have}\ "card\ S\ =\ card\ (\isasymUnion x\isasymin G.\ if\ x\ \isasymin \ S\ then\ \{x\}\ else\ \{\})"\isanewline
\ \ \ \ \ \ \isakeywordONE{by}\ simp\isanewline
\ \ \ \ \isakeywordONE{also}\ \isakeywordONE{have}\ "\isasymdots \ =\ (\isasymSum i\isasymin G.\ card\ (if\ i\ \isasymin \ S\ then\ \{i\}\ else\ \{\}))"\isanewline
\ \ \ \ \ \ \isakeywordONE{by}\ (intro\ card\_UN\_disjoint)\ (auto\ simp:\ finG)\isanewline
\ \ \ \ \isakeywordONE{also}\ \isakeywordONE{have}\ "\isasymdots \ =\ (\isasymSum x\isasymin G.\ if\ x\ \isasymin \ S\ then\ 1\ else\ 0)"\isanewline
\ \ \ \ \ \ \isakeywordONE{by}\ (force\ intro:\ sum.cong)\isanewline
\ \ \ \ \isakeywordONE{finally}\ \isakeywordONE{have}\ C:\ "card\ S\ =\ (\isasymSum x\isasymin G.\ if\ x\ \isasymin \ S\ then\ 1\ else\ 0)"\ \isakeywordONE{.}\isanewline
\ \ \ \ \isakeywordTHREE{show}\ ?case\isanewline
\ \ \ \ \ \ \isakeywordONE{using}\ insert\ \isakeywordONE{by}\ (auto\ simp:\ sum.distrib\ B\ C)\isanewline
\ \ \isakeywordONE{qed}
\end{isabelle}

We massage this into a version of equation~(\ref{eqn:F2}):
\begin{isabelle}
\ \ \isakeywordONE{have}\ "1/2\ \isasymle \ (sum\ card\ \isasymF )\ /\ (card\ \isasymF \ *\ card\ G)"\isanewline
\ \ \ \ \isakeywordONE{using}\ mult\_right\_mono\ [OF\ average\_ge,\ of\ "1\ /\ card\ G"]\ \isanewline
\ \ \ \ \isakeywordONE{using}\ cardG\_gt0\ \isakeywordONE{by}\ (simp\ add:\ divide\_simps\ split:\ if\_splits)\isanewline
\ \ \isakeywordONE{also}\ \isakeywordONE{have}\ "\isasymdots \ =\ (\isasymSum x\isasymin G.\ ((card\ \{S\isasymin \isasymF .\ x\isasymin S\})\ /\ (card\ \ \isasymF )))\ /\ card\ G"\isanewline
\ \ \ \ \isakeywordONE{by}\ (simp\ add:\ *\ sum\_divide\_distrib)\isanewline
\ \ \isakeywordONE{finally}\ \isakeywordONE{have}\ **:\ "1/2\ \isasymle \ (\isasymSum x\isasymin G.\ card\ \{S\isasymin \isasymF .\ x\isasymin S\}\ /\ card\ \isasymF )\ /\ card\ G"\ \isakeywordONE{.}
\end{isabelle}

Finally we set up the headline result as a proof by contradiction:
\begin{isabelle}
\ \ \isakeywordTHREE{show}\ ?thesis\isanewline
\ \ \isakeywordONE{proof}\ (rule\ ccontr)\isanewline
\ \ \ \ \isakeywordTHREE{assume}\ "\isasymnot \ union\_closed\_conjecture\_property\ \isasymF "\isanewline
\ \ \ \ \isakeywordONE{then}\ \isakeywordONE{have}\ A:\ "\isasymAnd \isasymX \ x.\ \isasymlbrakk \isasymX \isasymsubseteq \isasymF ;\ x\isasymin G;\ x\ \isasymin \ \isasymInter \isasymX \isasymrbrakk \ \isasymLongrightarrow \ card\ \isasymX \ <\ card\ \isasymF \ /\ 2"\isanewline
\ \ \ \ \ \ \isakeywordONE{by}\ (fastforce\ simp:\ union\_closed\_conjecture\_property\_def)
\end{isabelle}

Skipping a little massaging of the summation, we quickly reach the required contradiction.
\begin{isabelle}
\ \ \ \ \isakeywordONE{have}\ B:\ "(\isasymSum x\isasymin G.\ real\ (card\ \{S\isasymin \isasymF .\ x\isasymin S\}))\ <\ card\ \isasymF \ *\ (card\ G\ /\ 2)"\isanewline
\ \ \ \ \ \ \omitted\isanewline
\ \ \ \ \isakeywordONE{have}\ "(\isasymSum x\isasymin G.\ card\ \{S\isasymin \isasymF .\ x\isasymin S\}\ /\ card\ \isasymF )\ /\ card\ G\ <\ 1/2"\isanewline
\ \ \ \ \ \ \isakeywordONE{using}\ cardG\_gt0\ divide\_strict\_right\_mono\ [OF\ B,\ of\ "card\ \isasymF \ *\ card\ G"]\isanewline
\ \ \ \ \ \ \isakeywordONE{by}\ (simp\ add:\ divide\_simps\ sum\_divide\_distrib)\isanewline
\ \ \ \ \isakeywordONE{with}\ **\ \isakeywordTHREE{show}\ False\isanewline
\ \ \ \ \ \ \isakeywordONE{by}\ argo\isanewline
\ \ \isakeywordONE{qed}\isanewline
\isakeywordONE{qed}%
\end{isabelle}

Key aspects of the reasoning should be clear in such formal proofs, 
since at each point we write out what we are trying to prove. 
The reader never has to guess what is being proved.

Moreover, sometimes we need to express a formula in a particular way.
For example, suppose that
to derive a formula $a$ using a certain theorem we must first write it as an equivalent formula $a'$;
we can directly just do that. 
The alternative of massaging the subgoal $a$ into $a'$ using tactics
is a chore sometimes likened to manoeuvring cooked spaghetti with a spoon.
It's easier to prove two formulas to be equivalent than to transform one into the other.

\section{Conclusions} 

A common metric for a formalisation effort is its de Bruijn factor~\cite{wiedijk-de-bruijn}: 
the ratio of the size of the formal proof over that of the original. 
To minimise the distorting effect of file formats, blank lines, etc., 
it is typical to extract the original paper as plain text and compare the two files after compression. 
In our case, the Isabelle/HOL version is slightly smaller than the original: 
its de Bruijn factor is actually less than~1.
This is a remarkable outcome: a de Bruijn factor of 4 is more typical, and 10 is not unusual. 
To be sure, the two documents are not strictly comparable: 
the paper includes a literature review and bibliography,
while our development proves the material in Sect.\ts\ref{sec:uc-holds} (the counting argument~(5) and the proof by contradiction~(6)),
which is not explicitly shown in the original paper.
That we had to devote about 53 (out of the total 240) lines of code to proofs that were not in the paper 
and yet still ended up with a de Bruijn factor of less than~1, makes this observation even more interesting. 
The very low de Bruijn factor can be attributed to (a) not needing to introduce preliminaries from sumset theory as these had been already developed \cite{Pluennecke_Ruzsa_Inequality-AFP} and (b)
good automation powered by Sledgehammer that helped with efficient proof engineering.

Mathematicians will always be seeking understanding, and formalising mathematics
should be at the service of assisting this purpose. To this end, readable, well-organised code can prove to be useful: The value of this short formalisation lies not only in building up the Archive of Formal Proofs with a recent result in an active, modern area of research, but also, 
as any piece of formalised work should, in clarifying the proof. 
Readers will benefit from the high level of detail provided to fill in implicit arguments.
And we, the formalisers, through the process of formalisation, and often thanks to its interactive nature too, by rethinking definitions, notation and proof steps,
were thus able to gain, and hopefully to convey here, a better understanding of the proof.

\begin{credits}
\subsubsection{\ackname} 
The first author would like to thank David Ellis for informing her about the later related work \cite{Chase_Lovett}.
\end{credits}
%
%
%
\bibliographystyle{splncs04}
\bibliography{additive}

\end{document}